# A system for the deterministic transfer of 2D materials under inert environmental conditions

Patricia Gant[1,+], Felix Carrascoso[1,+], Qinghua Zhao[1,2,3], Yu Kyoung Ryu[1], Michael Seitz[4], Ferry Prins[4], Riccardo Frisenda[1,*], Andres Castellanos-Gomez[1,*]

[1] Materials Science Factory, Instituto de Ciencia de Materiales de Madrid, Consejo Superior de Investigaciones Científicas, 28049 Madrid, Spain
[2] State Key Laboratory of Solidification Processing, Northwestern Polytechnical University, Xi'an, 710072, P. R. China.
[3] Key Laboratory of Radiation Detection Materials and Devices, Ministry of Industry and Information Technology, Xi'an, 710072, P. R. China
[4] Condensed Matter Physics Center (IFIMAC), Autonomous University of Madrid, 28049 Madrid, Spain

[+] These authors contributed equally to this work

riccardo.frisenda@csic.es, andres.castellanos@csic.es

ABSTRACT

The isolation of air-sensitive two-dimensional (2D) materials and the race to achieve a better control of the interfaces in van der Waals heterostructures has pushed the scientific community towards the development of experimental setups that allow to exfoliate and transfer 2D materials under inert atmospheric conditions. These systems are typically based on over pressurized $N_2$ of Ar gloveboxes that require the use of very thick gloves to operate within the chamber or the implementation of several motorized micro-manipulators. Here, we set up a deterministic transfer system for 2D materials within a gloveless anaerobic chamber. Unlike other setups based on over-pressurized gloveboxes, in our system the operator can manipulate the 2D materials within the chamber with bare hands. This experimental setup allows us to exfoliate 2D materials and to deterministically place them at a desired location with accuracy in a controlled $O_2$-free and very low humidity (<2% RH) atmosphere. We illustrate the potential of this system to work with air-sensitive 2D materials by comparing the stability of black phosphorus and perovskite flakes inside and outside the anaerobic chamber.

The deterministic transfer methods, that allow for the placement of 2D materials onto a user defined specific location with an unprecedented degree of accuracy and reliability, are at the origin of the large success of two-dimensional (2D) materials research.[1–6] This control over the position of the transferred flakes has been exploited to fabricate devices with rather complex architectures as well as to assemble artificial stacks of dissimilar 2D materials to build-up the so-called van der Waals heterostructures.[7–10] In particular, deterministic placement methods have enabled the fabrication of

2D-based samples fully-encapsulated between insulating layers of hexagonal boron nitride (hBN), exhibiting record-high electronic performance that allowed for the observation of intriguing physical phenomena.[4,6] Nonetheless, there are two main challenges of this line of research: 1) how to achieve a better control of the interfaces,[11–13] 2) how to handle air-sensitive 2D materials.[14,15] Indeed, the continuous race to increase the performances of 2D devices and the isolation of novel unstable 2D materials under atmospheric exposure, have pushed towards the development of experimental setups where the transfer process can be carried out under controlled environmental conditions.[14,16,17] Although several research groups have implemented their own systems based on $N_2$ or Ar glove-boxes, the literature is lacking of comprehensive technical articles providing enough details for reproducing such systems by other groups. Moreover, these glove-box-based systems can be difficult to operate as delicate manual operation (e.g. mechanical exfoliation, handling small substrates with tweezers, …) have to be carried out while wearing thick gloves. In order to overcome this handicap, some systems are fully motorized and thus do not require any manual operation within the glovebox, increasing substantially their cost and complexity of implementation.[18]

Here, we present an alternative concept of deterministic transfer setup operated in inert conditions that is based on a gloveless anaerobic chamber. The user can operate inside the chamber using his/her bare hands without the need of thick glove-box gloves, making the transfer process virtually as easy as that carried out in air (outside the anaerobic chamber). Importantly, in the manuscript we provide all the details needed to facilitate the implementation of this system by other research groups.

Table 1 summarizes the different components needed to assemble our experimental system; whose cost is under the 20 000 €. We address the reader to the Supp. Info. for pictures of the different steps during the setup assembly process and for more details about the setup construction.

| Component | Description | Part number | Distributor | Price |
|---|---|---|---|---|
| Chamber * | Bactronez 12.5 CU.FT.I 354 L, anaerobic chamber, 300 plate capacity | BAAEZ22 | Bactron-Sheldon Manufacturing | 14700.00 € |
| Optomechanical component | Breadboard, ferromagnetic steel | 1BS-2040-015 | Standa | 130.00 € |
| | Post ¾" Dia Stainless Posts, Vertical Post 18" | 39-353 | Edmund Optics | 57.00 € |
| | Mounting Post Base, Ø2.48" x 0.40" Thick | PB1 | Thorlabs | 22.44 € |
| Zoom lens | Focusing, Coarse/Fine Movement, 50mm Diameter Thru Hole, Rack & Pinion | 54-792 | Edmund Optics | 355.00 € |
| | EOS Camera Adapter | 89-862 | Edmund Optics | 75.00 € |
| | 3.0X Mini-Camera Tube | 89-877 | Edmund Optics | 725.00 € |
| | 7X Zoom Module, Manual | 89-878 | Edmund Optics | 750.00 € |
| | Lower Module w/In-Line Illumination | 89-888 | Edmund Optics | 275.00 € |
| | 4.0X Lower Lens | 89-903 | Edmund Optics | 275.00 € |
| | Fiber Optic Light Guide Adapter | 89-919 | Edmund Optics | 125.00 € |
| | Fiber Optic Light Guide Adapter to 10mm | 89-920 | Edmund Optics | 55.00 € |
| Illumination | Bench Power Supply, Linear, Adjustable, 1 Output, 0 V, 30 V, 0 A, 5 A | 72-2690 | Farnell | 71.34 € |
| | LED, White, Through Hole, T-3 (10mm), 20 mA, 3.5 V, 8 cd | VAOL-10GWY4 | Farnell | 0.47 € |
| | Quartz Tungsten-Halogen Lamp | QTH10 | Thorlabs | 157.58 € |
| Oxygen meter | Oxygen detector | AR8100 | Smart Sensor | 81.10 € |
| Stages | Manual XY Linear Stage | MAXY-B60L-13 | Optics Focus Solutions | 72.10 € |
| | Manual XYZ Stage | MAXYZ-60L | Optics Focus Solutions | 164.00 € |
| | Square neodymiun magnets N52 10 x 10 x 4 mm | Amazon shop | Magnetastico | 18.99 € |
| | Ø1.40" Manual Mini-Series Rotation Stage, Metric | MSRP01/M | Thorlabs | 66.53 € |
| Camera | Canon EOS 1300D - 18 Mp (3" screen, Full HD, NFC, WiFi) | | Canon | 324.00 € |
| | Power supply for Canon EOS 1300D- ca. 3m, (ACK-E10) | 916606 | Subtel | 14.95 € |
| | Memory card | SDSDUNC-032G-GZFIN | SanDisk | 9.99 € |
| | Mini HDMI to HDMI cable | | AmazonBasics | 6.00 € |
| Screen | TV 32" Led HD | K32DLM7H | TD Systems | 139.00 € |
| TOTAL | | | | 18670.49 € |

**Table 1.** List of all the setup components, which includes the part numbers and the prices. * Note that the Bactronez anaerobic chamber includes: sleeves, elastic rubber sealing rings, sleeve compartment doors, integrated dry vacuum pump for the sleeve and the interlock purging, external lights, two sets of oxygen scrubber catalysts.

We build our transfer system within a gloveless anaerobic chamber (Bactronez 12.5). In order to access the interior part of the chamber, the system has two sealed doors connected to two plastic sleeves equipped with an elastic rubber sealing ring at their free end. The operator inserts his/her bare arms in the sleeves until feeling tight the rubber sealing ring around the arms like shown in Figure

1(b). Then, the sleeves are pumped down and purged with $N_2$ three times. After this process, the operator can open the two sealed doors without altering the environmental conditions inside the chamber. The system is equipped with a scrubber cartridge with Pd coated pellets (which act as a catalyst) that actively remove $O_2$ from the chamber through the reaction $2H_2+O_2 \xrightarrow{Pd} 2H_2O$[19] when the system is purged with forming gas ($N_2$ + 5% $H_2$) mixture. The $H_2O$ generated during the $O_2$ capture reaction is condensed in a big silica gel reservoir. We show in Figure 1(c) a comparison of the $O_2$ concentration evolution when the chamber is simply purged with $N_2$ ($O_2$ scrubber system OFF) and when the catalytic $O_2$ capture is employed ($O_2$ scrubber system ON). We found that the $O_2$ scrubber speeds up the $O_2$ concentration drop by a factor of 5 with respect to the simple $N_2$ purge. In fact, these kind of gloveless anaerobic chambers with active $O_2$ removal have proven to be very effective to generate environmental conditions with very low $O_2$ levels (in the 10 ppm range).

Figure 2 shows how the operator is preparing a Gel-Film (WF x4 6.0 mil, by Gel-Pak®) stamp with mechanically exfoliated black phosphorus. These pictures illustrate how the operator is able to easily carry out all the required delicate manual tasks (handling of tweezers, cutting, peeling, etc.) as he/she does not need to wear thick glove-box gloves. Briefly, a bulk piece of black phosphorus (HQ Graphene) is exfoliated using Nitto SPV224 tape. The tape containing the exfoliated black phosphorus flakes is gently pressed against the surface of a rectangle piece of Gel-Film substrate that will be used as viscoelastic stamp for the deterministic transfer.[2] The stamp is secured overhanging in the edge of a microscope glass slide with Scotch tape (Magic tape) and the glass slide is fixed to a XYZ micro-manipulator stage, that will be used to move the stamp, with double side tape (Scotch restickable tabs). In order to identify the flake to be transferred, we place a collimated light source (Thorlabs quartz tungsten-halogen lamp, QTH11) under the stamp allowing us to infer the thickness of the flakes from its transmittance.[20] Once the desired flake is located, the lamp is replaced by an XY-rotation stage that will be used to move the acceptor sample (also fixed to the stage with double side Scotch restickable tabs).

Figure 3 shows a sequence of optical microscopy images acquired with the transfer system zoom lens during the deterministic placement of a black phosphorus flake onto a $SiO_2$/Si substrate with a pre-patterned alignment crosshair. First, the flake is aligned in the center of the crosshair markers. Then, the Gel-Film stamp is lowered until establishing a gentle contact between stamp and acceptor sample. Finally, the stamp is slowly peeled off until it is completely removed, and the flake is successfully transferred in the middle of the crosshair. The possibility to place flakes at a desired location (as shown in Figure 3) can be further exploited to fabricate van der Waals heterostructures by stacking different 2D materials on top of each other. Figure 4 demonstrates the capability of our inert atmosphere deterministic placement system to fabricate this kind of artificial stacks of 2D materials. The different transfer steps to fabricate a heterostructure based on a few-layer black phosphorus (BP) flake sandwiched between two flakes of hexagonal boron nitride (hBN) are shown in Figure 4.

In order to benchmark the potential of the gloveless anaerobic chamber to manipulate air-sensitive 2D materials, we monitor the evolution of the black phosphorus flake (transferred in Figure 3) inside the anaerobic chamber through optical microscopy images with the system's zoom lens. Black phosphorus is well-known for its instability under air exposure.[21–23] We do not notice any change in the optical microscopy images after 22 days inside the chamber. On the other hand, after taking the flake outside the chamber, it quickly shows signs of degradation from day 3 of air exposure (see inset in Figure 5 with increased contrast) and it becomes severely damaged after 9 days of exposure. In Figure 5, we show the optical contrast profiles acquired from the optical images of the black phosphorus sample along the black line indicated in the first panel. The profiles present a clear increase in roughness under air exposure due to the degradation, comparing with the profiles of the images of the flake inside the chamber. This illustrates the potential of this experimental setup for handling air-sensitive 2D materials.

Furthermore, we also checked the performance of the gloveless anaerobic chamber with even more unstable 2D materials. We studied a compound that belongs to the metal halide perovskites family that is known to quickly degrade upon atmospheric exposure.[24–26] The metal halide perovskites are a prominent group of materials in the solar harvesting research,[27–29] optoelectronics[30–32] and photovoltaics[33,34] which degrade rapidly when exposed to air or light.[24,35,36] In our case, the compound used is phenethylammonium lead iodide ($PEA_2PbI_4$) 2D perovskites synthesized according to the techniques described in References.[26,37,38] The bulk crystals are exfoliated and transferred to a $SiO_2$/Si substrate inside the anaerobic chamber. Then, the sample is monitored for 5 hours by acquiring microscope optical pictures, similar to the procedure used in the black phosphorus sample (Figure 6). Again, no apparent change of the material is observed during the 5 hours in inert atmosphere. However, the surface of the perovskite flakes quickly showed degradation signs after just 15 minutes outside the chamber. In this case, we show the optical contrast of the flake to illustrate the degradation upon air exposure (see the Supporting Information for the analysis of the time evolution of the optical contrast). Given the results, we conclude that the atmosphere in the anaerobic chamber allows the handling of very air-sensitive 2D materials.

In summary, we have developed a system to precisely transfer 2D materials under inert atmosphere and we provide all necessary technical details to allow its implementation in other laboratories. Unlike other inert atmosphere transfer system used in the literature based on $N_2$ or Ar over pressurized gloveboxes, our system relies on a gloveless anaerobic chamber with an active $O_2$ catalytic capturing system. Our setup allows the operator to work with his/her bare hands within the chamber, which facilitate tasks as scissors cutting, tape peeling, tweezer handling, etc. that are challenging to be done while wearing thick gloves required to access the interior of an over pressurized $N_2$ or Ar glove-box. We illustrate the operation of our transfer setup by showing the deterministic transfer of a black phosphorus flake and by studying the stability of thin black phosphorus and perovskite flakes. We believe that the simple operation and the low cost (<20 000 €) of this setup makes it highly attractive for other research groups that have not yet implemented a transfer system under inert atmosphere.

**Acknowledgements**

This project has received funding from the European Research Council (ERC) under the European Union's Horizon 2020 research and innovation programme (grant agreement n° 755655, ERC-StG 2017 project 2D-TOPSENSE). EU Graphene Flagship funding (Grant Graphene Core 2, 785219) is acknowledged. R.F. acknowledges the support from the Spanish Ministry of Economy, Industry and Competitiveness through a Juan de la Cierva-formación fellowship 2017 FJCI-2017-32919. QHZ acknowledges the grant from China Scholarship Council (CSC) under No. 201700290035. MS acknowledges the financial support of a fellowship from "la Caixa" Foundation (ID 100010434). The fellowship code is LCF/BQ/IN17/11620040. MS has received funding from the European Union's Horizon 2020 research and innovation program under the Marie Skłodowska-Curie grant agreement No. 713673. FP acknowledges financial support from the Spanish Ministry of Economy and Competitiveness through the "María de Maeztu" Program for Units of Excellence in R and D (MDM-2014-0377).

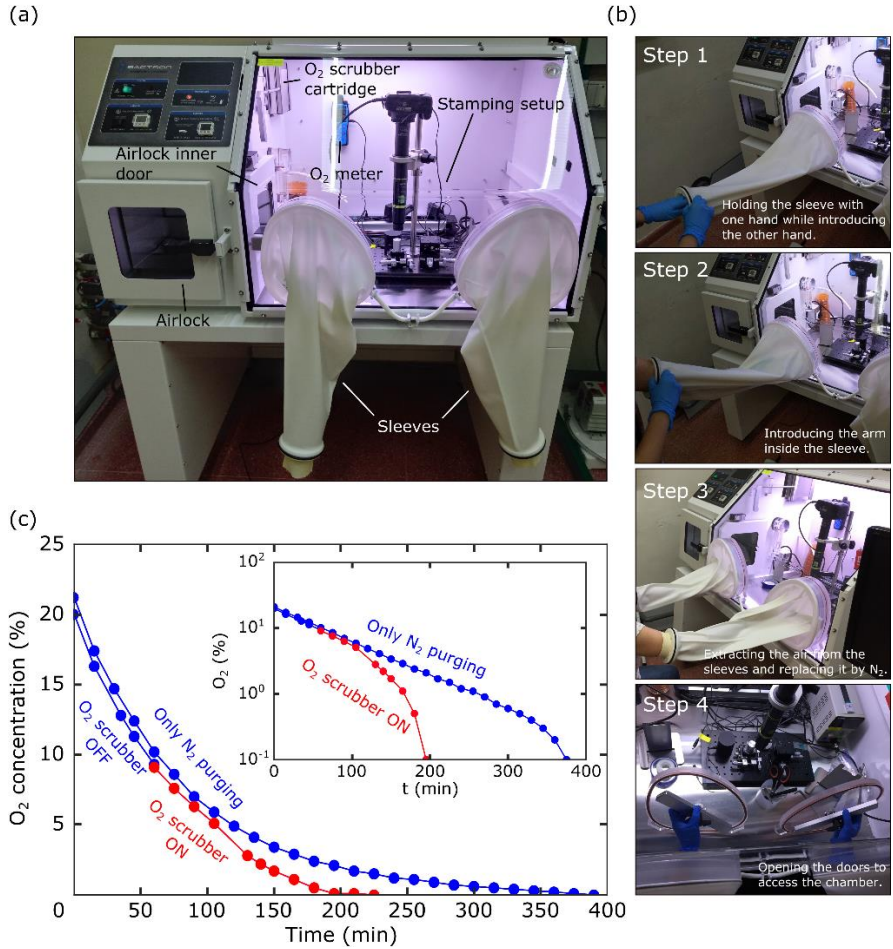

**Figure 1.** (a) Optical image of the gloveless chamber with the stamping setup mounted inside. (b) Optical images of the steps needed to introduce the hands inside the chamber. (c) Oxygen concentration drop in time within the chamber for a simple $N_2$ purging (blue dots line), and forming gas purging with the $O_2$ scrubber turned on after 50 minutes (blue and red dots line). Inset: Oxygen concentration shown in log scale.

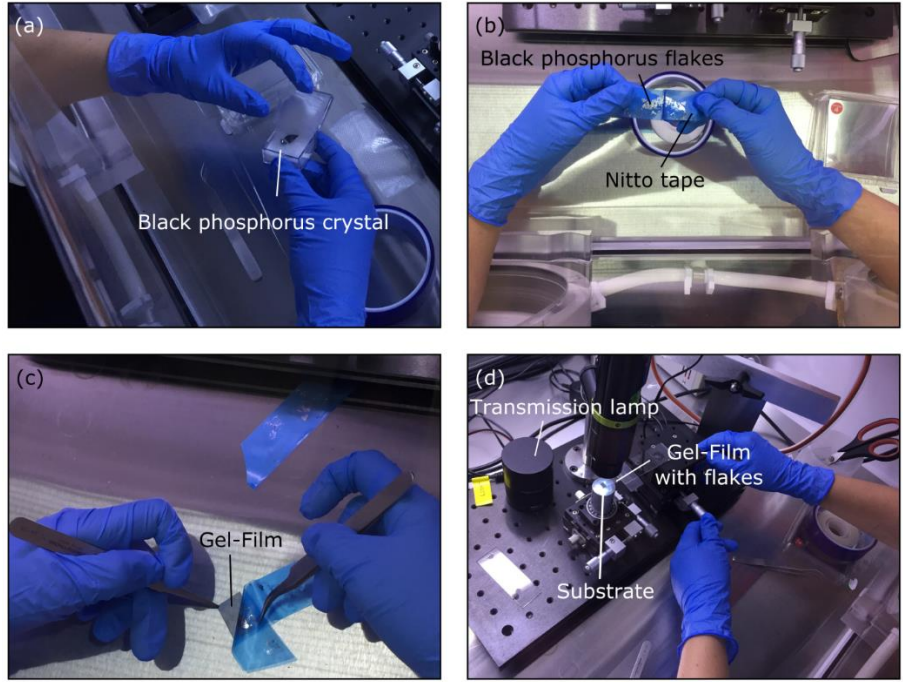

**Figure 2.** (a) Manipulation of the box with the bulk black phosphorus inside the chamber. (b) Mechanical exfoliation with Nitto tape by hand in the chamber. (c) Black phosphorus flakes being transferred onto the Gel-Film stamp after exfoliation with Nitto tape. (d) Handling the micrometer XYZ stage to deterministically transfer a flake.

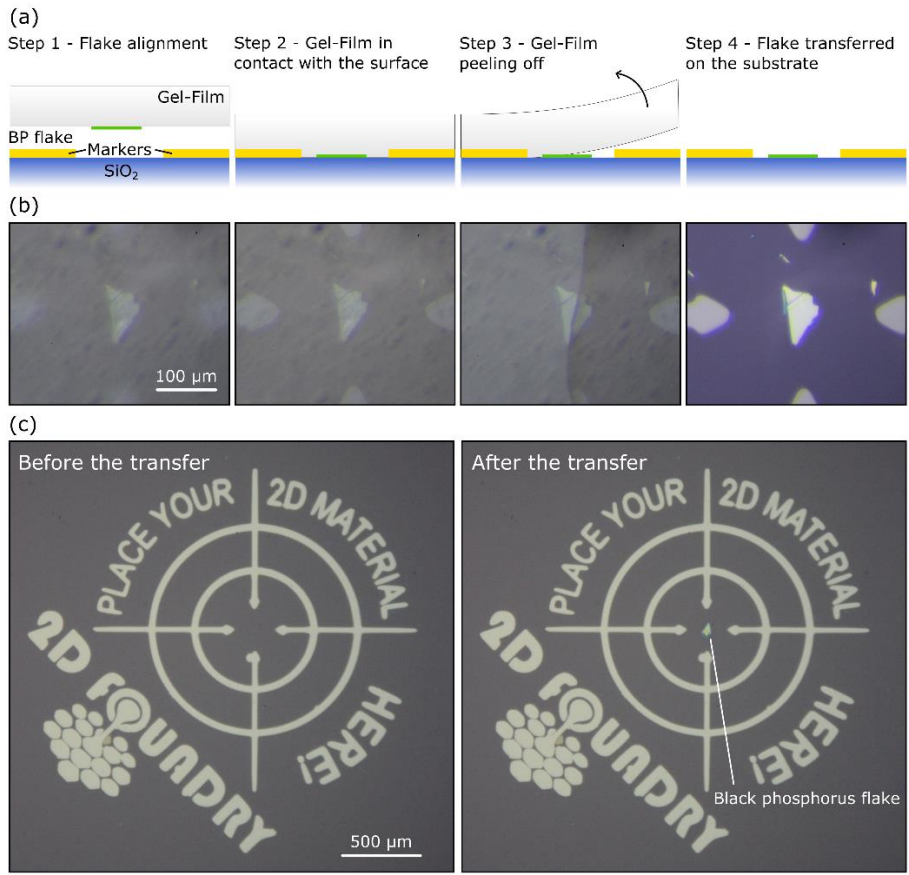

**Figure 3.** (a) Scheme of deterministic transfer process of black phosphorus flakes on marked substrates. (b) Optical images of the transfer of a black phosphorus flake in the gloveless chamber. (c) Optical images of the substrate before and after the transfer of the black phosphorus flake.

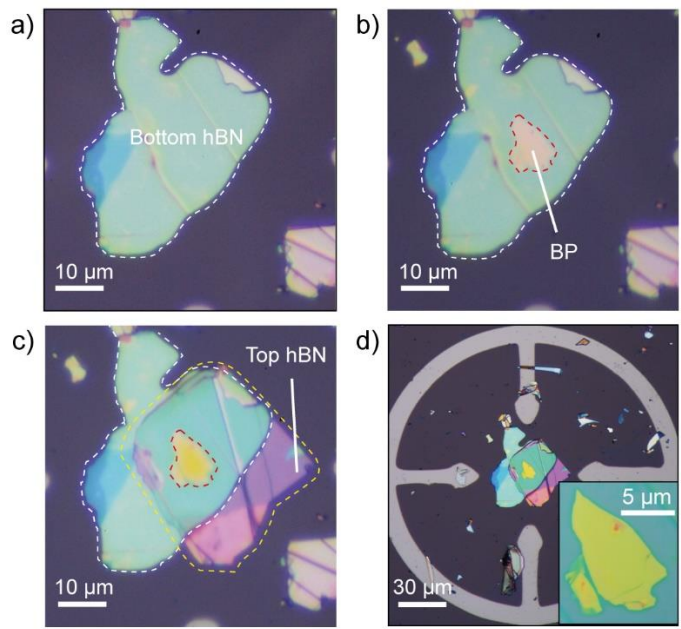

**Figure 4.** Fabrication of a van der Waals heterostructure under inert atmospheric conditions. (a) Optical image of hexagonal boron nitride (hBN) flake transferred onto a SiO$_2$/Si substrate. (b) black phosphorus (BP) flake transferred onto the boron nitride flake. (c) a top hBN flake is transferred to complete the hBN/BP/hBN stack. (d) a low magnification image of the assembled heterostructure in the middle of a crosshair marker. The inset shows a higher magnification optical microscopy image of the hBN/BP/hBN heterostructure.

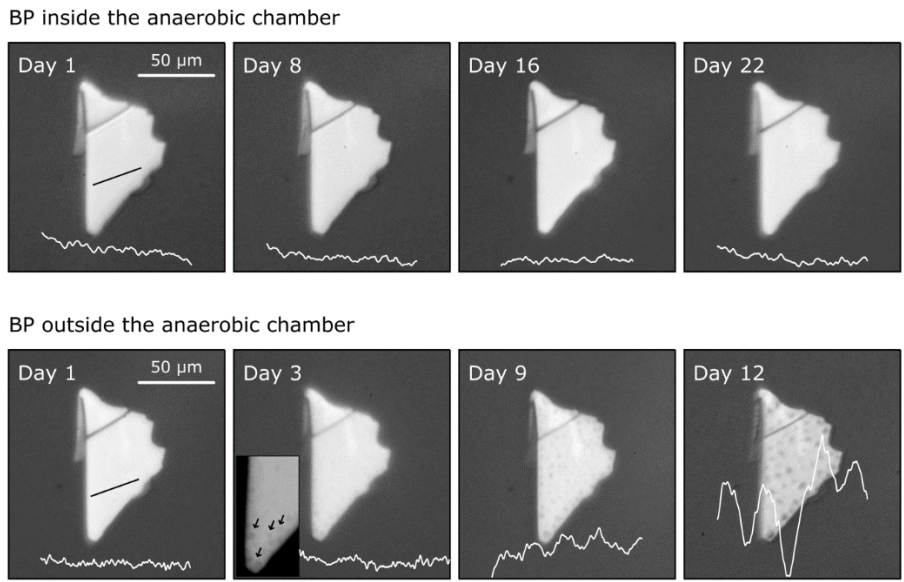

**Figure 5.** Optical images (showing the green channel which provides the highest contrast) of the same BP flake during several days inside the gloveless chamber (22 days) and outside it (12 days).

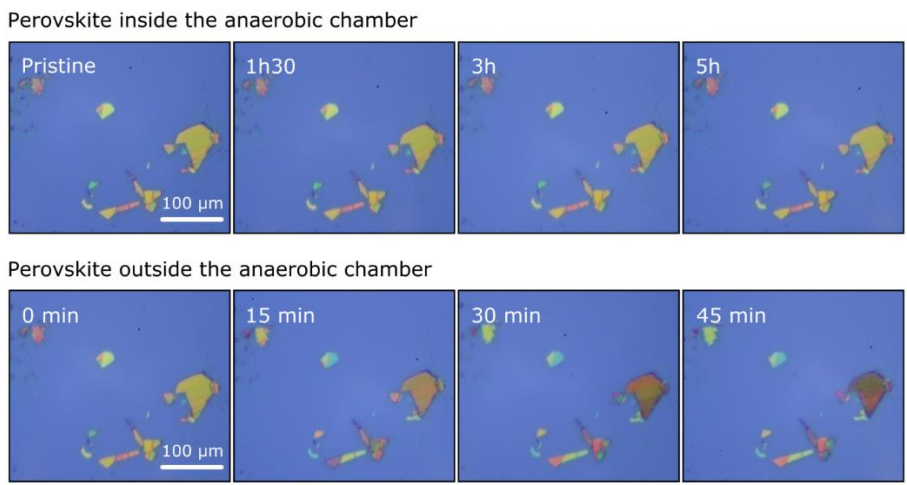

**Figure 6.** Optical images of the same perovskite flakes during several hours inside the gloveless chamber (5 h) and outside (45 min).

Supporting Information:

# A system for the deterministic transfer of 2D materials under inert environmental conditions

Patricia Gant[1,+], Felix Carrascoso[1,+], Qinghua Zhao[1,2,3], Yu Kyoung Ryu[1], Michael Seitz[4], Ferry Prins[4], Riccardo Frisenda[1,*], Andres Castellanos-Gomez[1,*]

[1] Materials Science Factory, Instituto de Ciencia de Materiales de Madrid, Consejo Superior de Investigaciones Científicas, 28049 Madrid, Spain
[2] State Key Laboratory of Solidification Processing, Northwestern Polytechnical University, Xi'an, 710072, P. R. China.
[3] Key Laboratory of Radiation Detection Materials and Devices, Ministry of Industry and Information Technology, Xi'an, 710072, P. R. China
[4] Condensed Matter Physics Center (IFIMAC), Autonomous University of Madrid, 28049 Madrid, Spain

[+] These authors contributed equally to this work

riccardo.frisenda@csic.es, andres.castellanos@csic.es

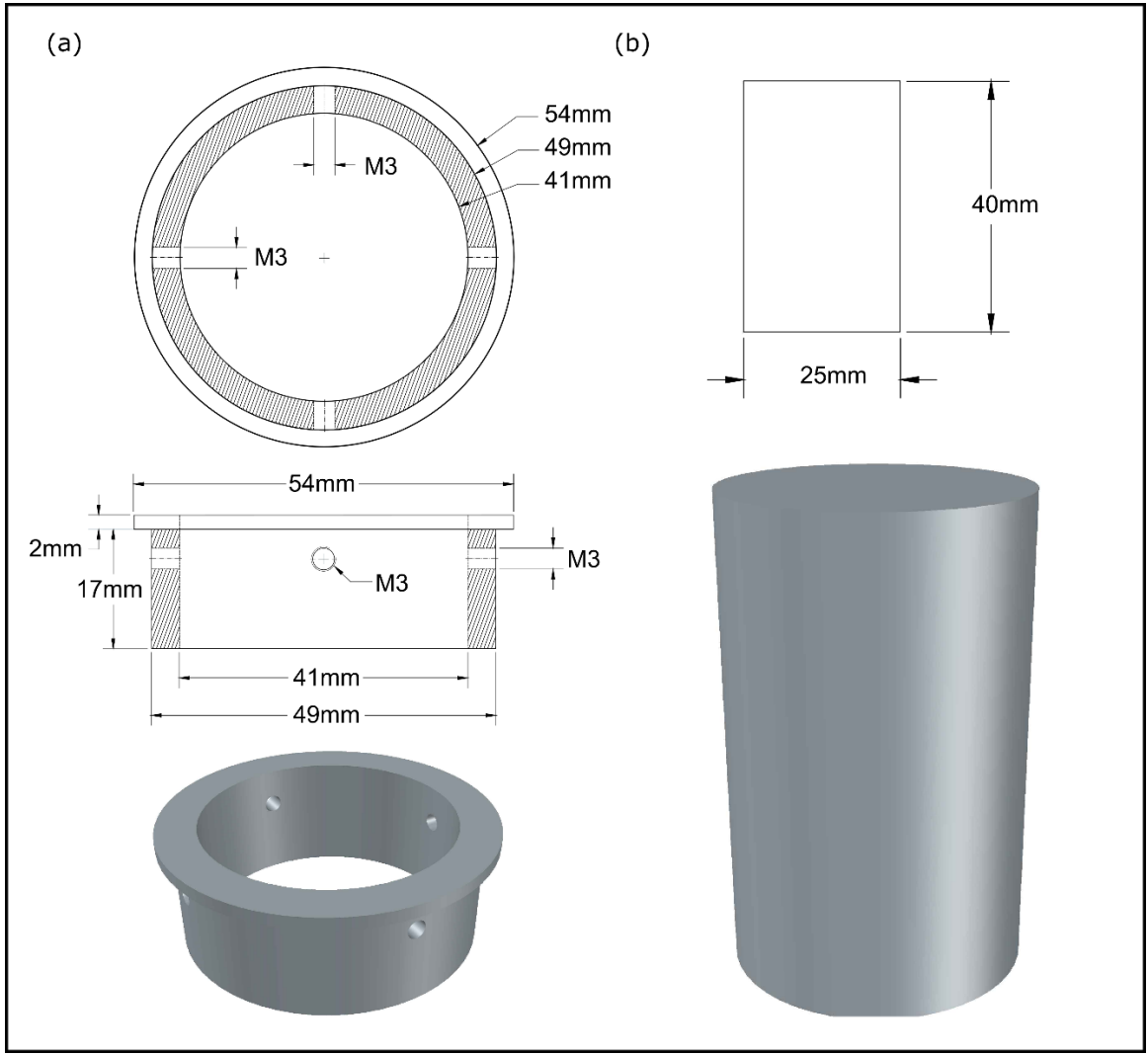

**Figure S1.** (a) Schematic of the home-made piece 1, which is designed for holding the camera tube to the focussing system. (b) Schematic of the home-made piece 2, which is attached to the rotator placed in the XY-microstage for adjusting the height of the sample.

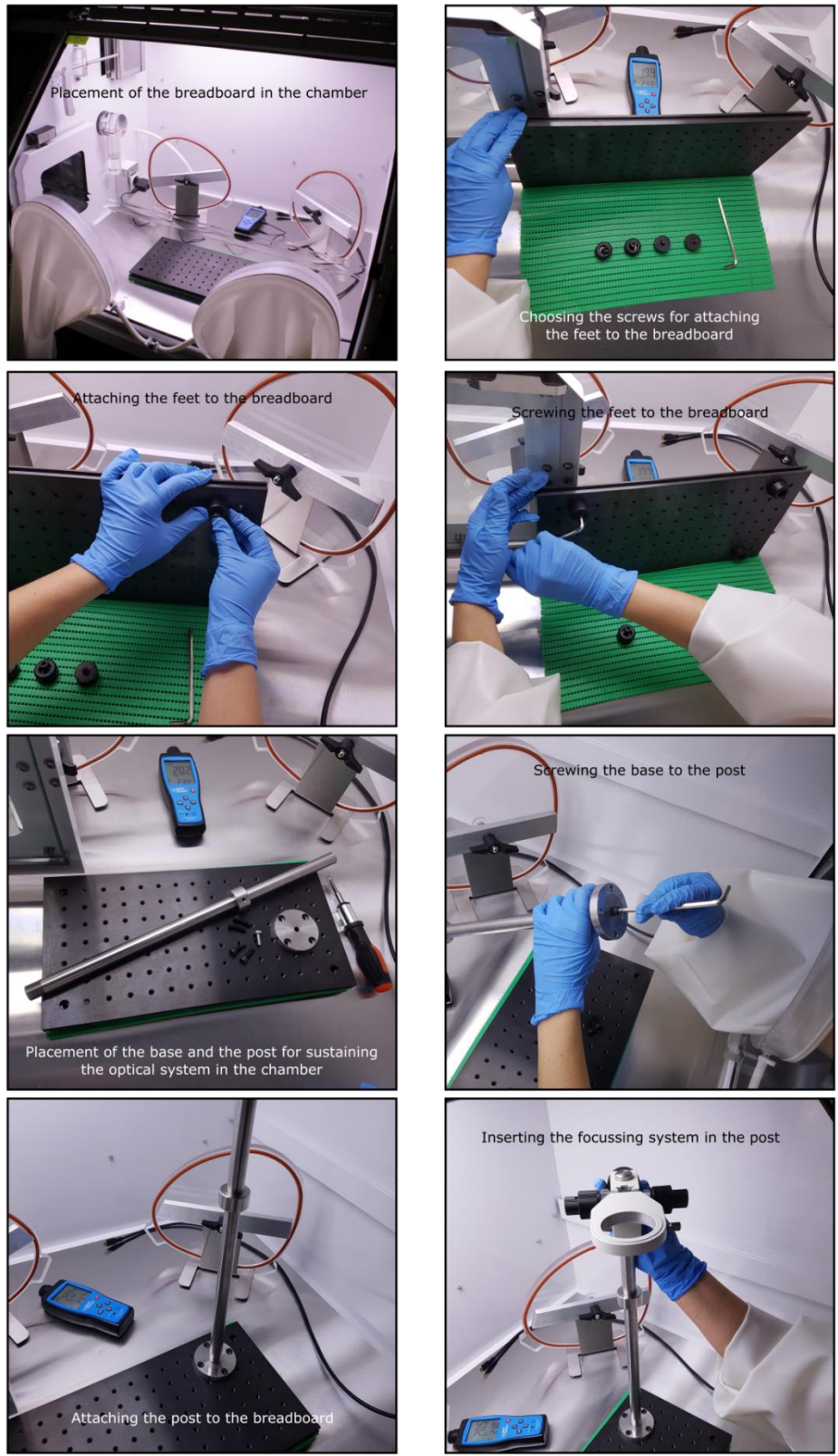

**Figure S2.** Optical images of the mounting process of the transfer setup inside the gloveless box chamber. In this part of the process, the breadboard, the post and the focusing system are set up in the chamber.

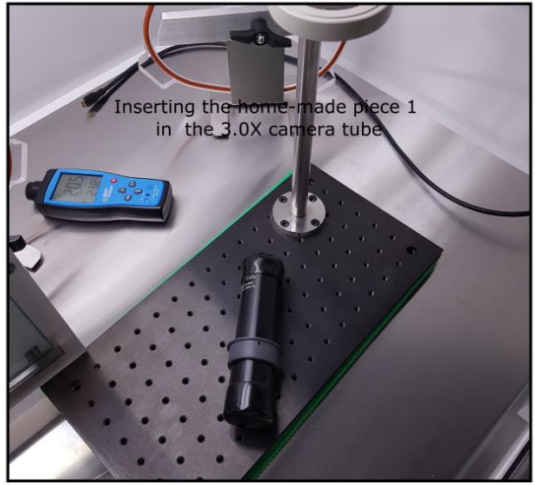
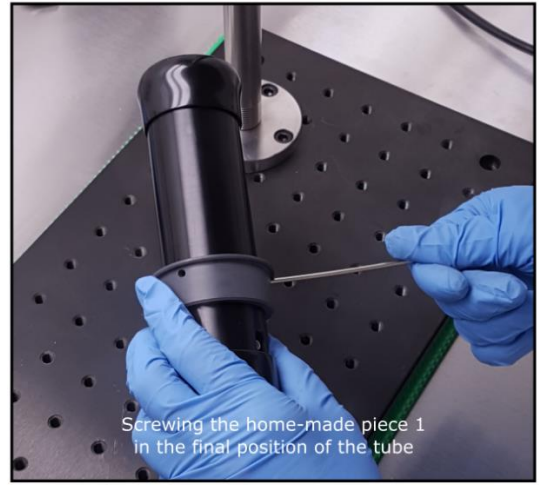
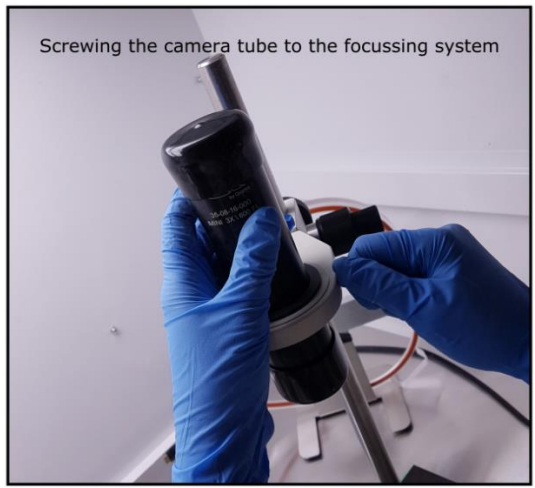
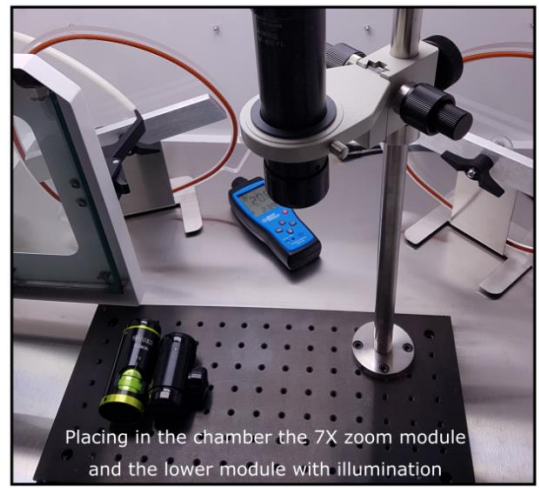
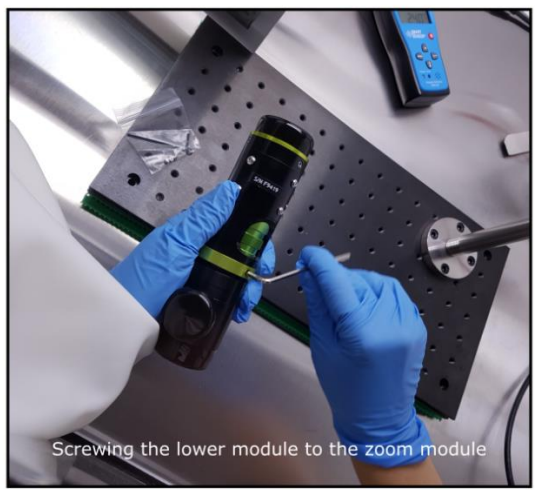
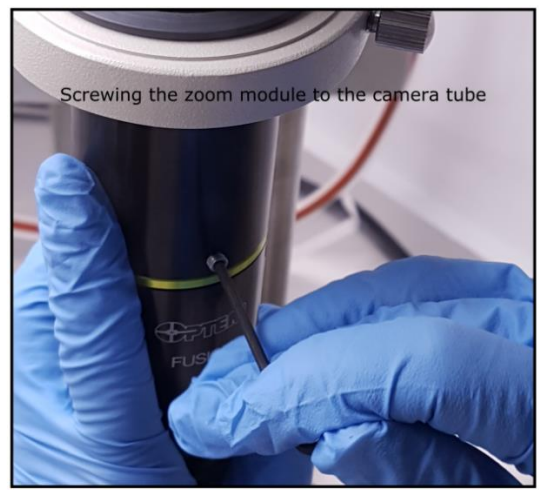

**Figure S3.** Optical images of the mounting process of the transfer setup inside the gloveless box chamber. In this part of the process, the main parts of the zoom lens are mounted in the focusing system.

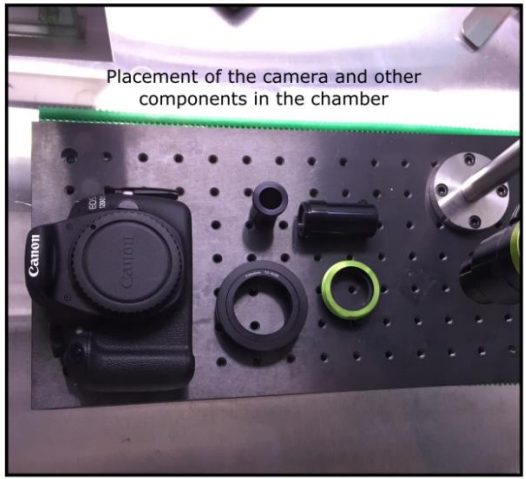
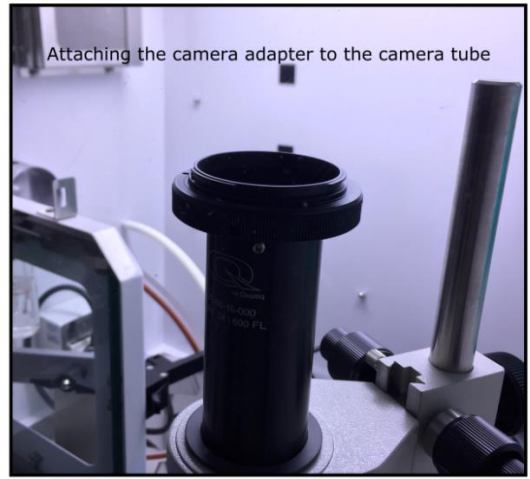
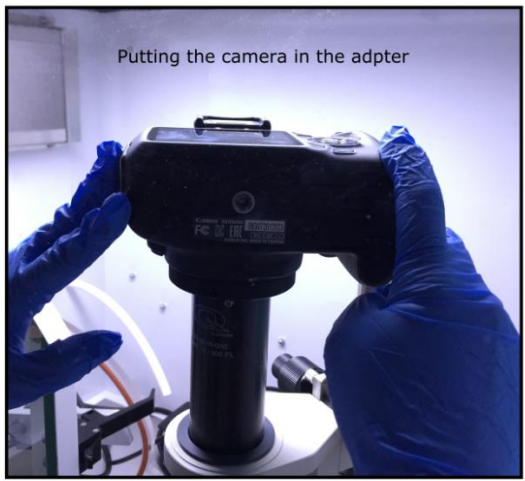
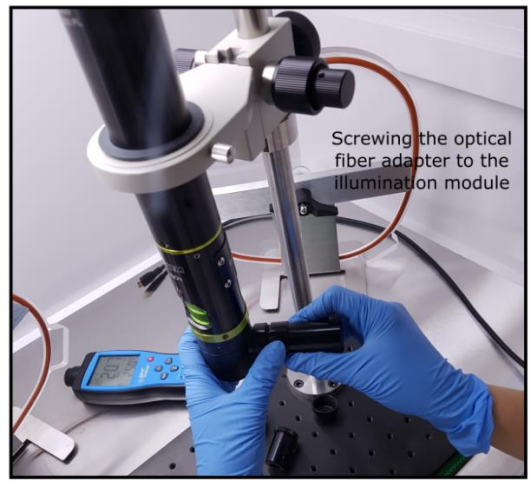
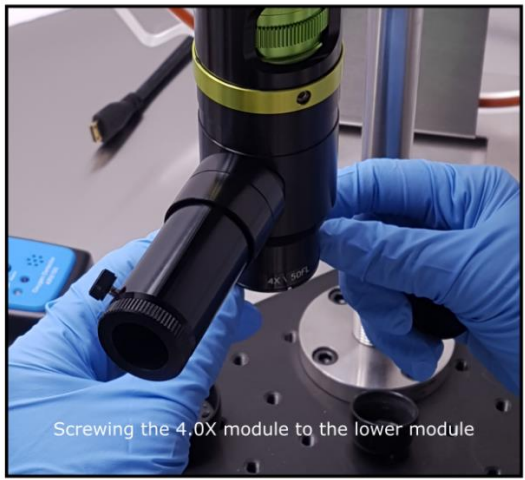
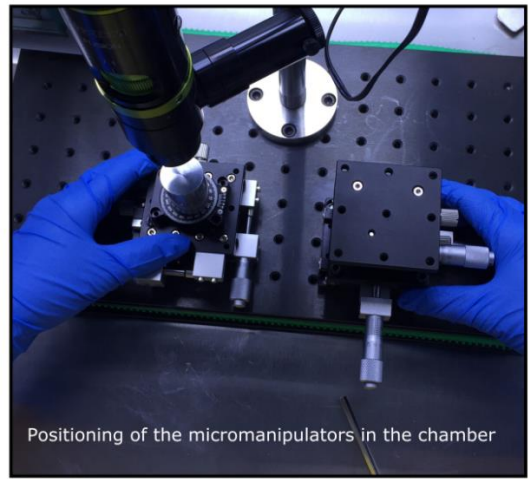

**Figure S4.** Optical images of the mounting process of the transfer setup inside the gloveless box chamber. In this part of the process, the camera, fiber adapter and an extra magnification module are mounted in the zoom lens. The micromanipulators are also introduced in the chamber

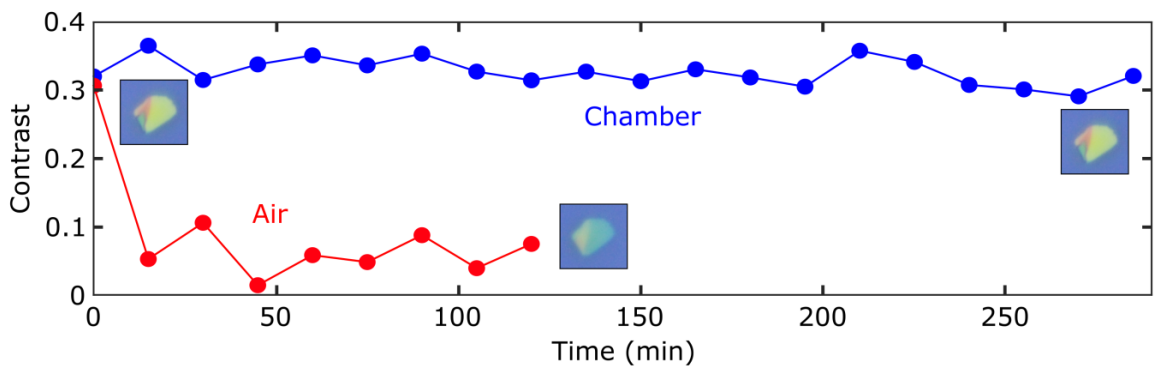

**Figure S5.** Optical contrast of the thin perovskite flake versus the time when the sample is inside the anaerobic chamber (blue) and outside the anaerobic chamber (red).

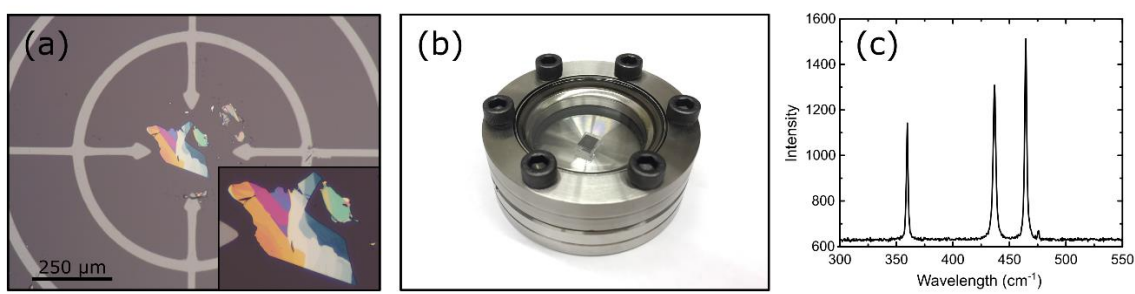

**Figure S6.** Preserving air-sensitive samples, fabricated within the anaerobic chamber, for their ex-situ optical characterization. (a) Optical image of a black phosphorus flake transferred within the anaerobic chamber on the middle of a target substrate with a pre-patterned cross-hair. (Inset) Zoom in image of the black phosphorus shown in (a). After the flake is transfer within the anaerobic chamber, the substrate is placed inside a miniature optical chamber that is naturally filled with the anaerobic atmosphere present in the gloveless chamber. (b) Photograph of the miniature chamber with a sample loaded inside. The miniature optical chamber can be unloaded from the anaerobic chamber and it can be placed in other spectroscopic setups. (c) Ex-situ Raman spectrum of the black phosphorus flake shown in (a) with the substrate placed in the miniature optical chamber.

| Optical testing chamber | | | | | |
|---|---|---|---|---|---|
| **Description** | **Part number and link** | **Distributor** | **Quantity** | **Unit price** | **Price** |
| Screws | M6X25NBW | Hositrad | 25 | 0.45 € | 11.25 € |
| Vacuum gasket | VG38 | Hositrad | 1 | 2.3 € | 2.3 € |
| Optical window | HVP-1500 | Hositrad | 1 | 85 € | 85 € |
| Blank flange | CF35/BT | Hositrad | 1 | 17 € | 17 € |
| | | | | **TOTAL** | **115.55 €** |

**Table S1.** List of all the components needed to assemble the miniature optical chamber for the ex-situ characterization of the samples fabricated within the anaerobic gloveless chamber.

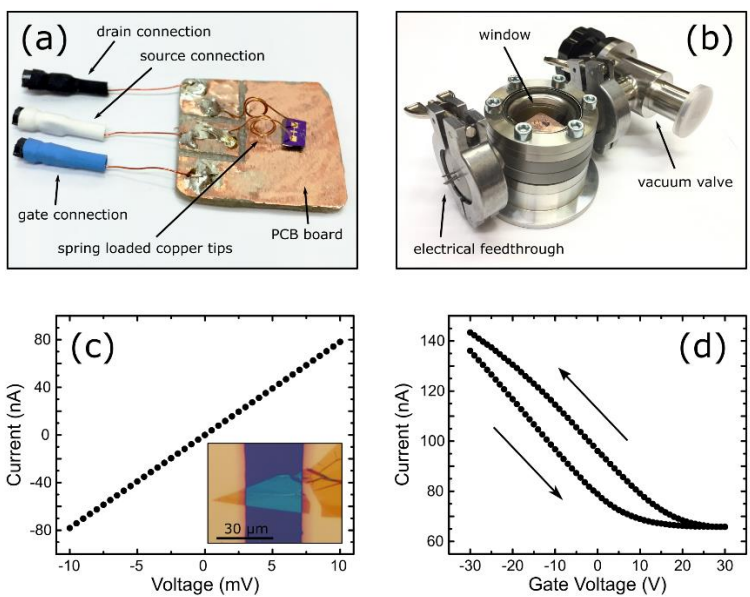

**Figure S7.** Preserving air-sensitive samples, fabricated within the anaerobic chamber, for their ex-situ electrical and optoelectronic characterization. Simple three-terminal electronic devices can be fabricated by using the deterministic transfer system to transfer a 2D material flake bridging two pre-patterned electrodes deposited on Si(p++)/SiO$_2$. (a) After the transfer is finished, the substrate can be loaded in a small printed-circuit board (PCB) equipped with two spring loaded copper tips that will make electrical connection to the source and drain electrodes. Electrical connection to the gate can be done by connecting the back part of the substrate with silver paint (using the heavily doped silicon substrate as a gate electrode). (b) The PCB board can be loaded in a miniature chamber equipped with electrical connections, an optical window and a vacuum valve. After closing the miniature chamber, the sample is preserved in an anaerobic atmosphere. The chamber can be evacuated through the vacuum valve to allow electrical transport measurement at high vacuum levels (10$^{-6}$ mbar). (c) and (d) Current *vs.* voltage characteristics ($V_g = 0$ V) of a black phosphorus field effect transistor fabricated within the anaerobic chamber (see inset for a microscope image of the device), recorded in vacuum, and current vs. gate voltage characteristics ($V_{sd} = 0.01$ V).

| Electrical testing chamber | | | | | |
|---|---|---|---|---|---|
| **Description** | **Part number and link** | **Distributor** | **Quantity** | **Unit price** | **Price** |
| Screws | M6X25NBW | Hositrad | 25 | 0.45 € | 11.25 € |
| Vacuum gasket | VG38 | Hositrad | 2 | 2.3 € | 4.6 € |
| Optical window | HVP-1500 | Hositrad | 1 | 85 € | 85 € |
| Blank flange | CF35/BT | Hositrad | 1 | 17 | 17 € |
| Double-sided flange with 2x KF25 ports | CF35/38DX2/25 | Hositrad | 1 | 75 € | 75 € |
| KF25 manual vacuum valve | VAV25KF | Hositrad | 1 | 235 € | 235 € |
| KF25 electrical feedthrough 3-pins | 17050-01-KF | Hositrad | 1 | 204 € | 204 € |
| KF25 centering ring | KF25/RA | Hositrad | 2 | 3 € | 6 € |
| KF25 Clamp | KF25/C | Hositrad | 2 | 3.75 € | 7.5 € |
| | | | | **TOTAL** | **645.35 €** |

**Table S2.** List of all the components needed to assemble the miniature optoelectronic chamber for the ex-situ electrical characterization of the samples fabricated within the anaerobic gloveless chamber.